\documentclass[10pt]{article}
\usepackage{url}

\usepackage{perpage}
\MakePerPage{footnote}

\usepackage{a4,amsmath,amssymb,graphicx}

\def\displayfrac#1#2{\frac{\displaystyle #1}{\displaystyle #2}}
\newcommand{\nx}{\noindent}

\begin{document}
\title{Einstein's Miraculous Year}

\author{Vasant Natarajan, V Balakrishnan and N Mukunda}

\maketitle

\begin{abstract}
With each passing year, the young Albert Einstein's achievements in physics in the year 1905 seem to be ever more miraculous. We describe why the centenary of this remarkable year is worthy of celebration.
\end{abstract}

\section*{Introduction}

The revolution of the earth around the sun has given us a
natural unit of time, the year. Since time immemorial, notable
events in human affairs have been marked out by
the {\it year} in which they occurred. Commemorations are customarily
held every
twenty-five years after the event. Of these, the {\it centenary}
is very special.
If the centenary of an event is celebrated, it signifies two
things: on the one hand,
a hundred years is a sufficiently long period to claim that
the importance of the event has stood the test of time; at the same
time, it is a period short enough to be almost within living memory,
so that the historical setting of the event
can be recalled reliably.

In Science, too, there have occurred many notable events and
discoveries that justify
centenary celebrations. But there are a select few that are
more than notable: they are {\it watershed} events for the human race
itself, in a far more profound sense
than mere political events (however tumultuous the latter may appear
to be when they occur). They separate distinct
eras in humankind's
understanding of the universe in which it lives. The year 1905 was,
without question, such a vintage year. The current year,
2005, marking the centenary of that remarkable year,
has been declared the International Year of
Physics by organizations such as the UN and UNESCO.  It is
being observed by special programmes, lectures and seminars in
a large number of countries, including India.

1905 was Albert Einstein's {\it Annus Mirabilis} or
`Miraculous Year'.
Between March and December that year,
the 26-year-old Einstein published six seminal papers in the journal
{\it Annalen der Physik}
that advanced --- indeed, revolutionized --- our understanding
of the physical universe in major ways in three different
directions. In the order in which they appeared,
the papers (see Box 1) dealt with (i) the `light-quantum' or the
photon concept and an explanation of the photoelectric effect, (ii) the
theory and explanation of Brownian motion, and (iii)
the Special Theory
of Relativity, a radically new view of space and time. Einstein himself
regarded the first as truly revolutionary; it was the second major
step in the development of quantum theory. In contrast,
both Brownian motion
and Special Relativity belong to the realm of classical physics.
In addition, in 1905, Einstein discovered
the equivalence of mass and energy, encapsulated in perhaps the most
famous formula of all: $E=mc^2$. No single year before or
since then has seen
such a diversity of fundamental discovery by a single person, with the
exception of the period 1665-66 in
which Isaac Newton, also in his early twenties,
discovered `the system of the world', and much else besides (see Box 2).

The previous decade in physics had seen three major experimental
discoveries in physics. X-rays were discovered
by Wilhelm Rontgen in 1895, in Germany;
Henri Becquerel in France discovered radio-activity
in 1896; and in England, J. J. Thomson identified the electron
in 1897. Shortly after that, in 1900, Max
Planck had taken the first step toward quantum theory with his
Law of Temperature Radiation. It is helpful to remember also
that, at that time, there still remained some prominent scientists
-- the physicist-philosopher Ernst Mach and the chemist
Wilhelm Ostwald among them --
who doubted the atomic nature of matter. Galaxies other than our own
were unknown, and it was thought that the Milky Way comprised the
entire universe. Powered flight of a heavier-than-air vehicle had just been
demonstrated by the Wright brothers in 1903.
Needless to say, most of the technological marvels we
take for granted today
(jet aircraft, mobile phones, satellite TV, computers) were completely
unknown.

To properly understand the significance of
Einstein's three major achievements of 1905, we have to set the stage
by going a bit further back in history.

\section*{The genesis of the photon}

In 1859 Gustav Kirchoff had posed the following problem: to measure
experimentally, and explain theoretically, the energy distribution
of `Temperature Radiation' over different frequencies of the
radiation.
If we have radiation in thermal equilibrium with material bodies at
a common absolute temperature $T$,
how much of its energy density lies in each
small range $(\nu\,,\,\nu+ d\nu)$ of frequencies?
In the years that followed, many
physicists --- Stefan, Boltzmann and Wien, among others ---
made important contributions toward the solution of
the problem. Wien not only proved
a theorem showing that energy density $\rho(\nu\,,\,T)$ must have the
form
     \[\rho(\nu\,,\,T) = \alpha\;\nu^3\;f(\nu/T),\]

     \nx
     but also suggested that the function $f$ had an exponential form,
     so that
     \[\rho(\nu,T)
= \alpha \,\nu^3\,e^{-b\nu/T}\quad (b = {\rm constant}).\]
     \nx
For a while, Planck believed that Wien's formula was exact, i. e.,
valid for all frequencies $\nu$, and made several unsuccessful attempts to
derive it from first principles.  In late 1900, however, he learnt
that the formula was in agreement with experimental observations
only for large $\nu$, and not for small $\nu$. At low frequencies
the experimental data agreed with the
Rayleigh-Jeans Law, according to which
$ \rho(\nu\,,\,T)  =(8\pi\nu^2/c^3)\,(k_{B}T)$, where
$k_B$ is Boltzmann's constant. This was
the unique form for $\rho$ predicted on the
basis of Maxwell's equations of
electromagnetism combined with classical statistical physics. Within
a few hours of learning of this
situation, he had found a formula for $\rho$ that interpolated
between these two frequency regimes:
%\begin{eqnarray*}
%\begin{array}{cccccl}
%&&\mbox{low}\;\nu&&&\displayfrac{8\pi\nu^2}{c^3}\cdot k_{B}T :
%\mbox{R-J}\nonumber\\
%\rho(\nu,T)=\displayfrac{8\pi\nu^2}{c^3}
%\left(\displayfrac{h\nu}{e^{h\nu/k_{B}T}
%-1}\right)&&&&&\nonumber\\
%&&\mbox{high}\;\nu&&&\displayfrac{8\pi\nu^2}{c^3}
%\cdot h\nu e^{-h\nu/k_{B}T}:\mbox{Wien}
%\end{array}
%\end{eqnarray*}
\begin{eqnarray*}
\rho(\nu\,,\,T)=
\displayfrac{8\pi\nu^2}{c^3}
\left(\displayfrac{h\nu}
{e^{h\nu/k_{B}T} -1}\right)
\longrightarrow
\left\{
\begin{array}{ll}
(8\pi\nu^2/c^3)\,(k_{B}T),
\quad & {\rm for \ low}\,\nu,\\[6pt]
(8\pi\nu^2/c^3)\,
(h\nu \,e^{-h\nu/k_{B}T})
\quad & {\rm for \ high}\,\nu.
\end{array}\right.
\end{eqnarray*}
The formula above (now known as
Planck's Law) involved a new constant of nature, $h$, now called
Planck's constant. It fit the data for all frequencies.
Over the next couple of months he constructed a mechanism, an
argument, that would lead to the formula. To do this he
made the assumption that
matter had only discrete, i. e., quantized, energy values, so that it
could only absorb and emit discrete amounts of radiative
energy. However, radiation itself was assumed to obey Maxwell's
equations exactly. Its energy could therefore vary continuously from
zero upward.

In 1905 Einstein presented an incisive analysis of Planck's
Law in the Wien or high-frequency limit,
which was known to be the non-classical
regime --- clearly, this was where something new could be learnt
about radiation.  He considered Wien radiation with energy $E$
at a frequency $\nu$ contained in a volume $V$, and found
the result (in modern notation)\\[6pt]
\nx
``$\cdots$ the probability that at a randomly chosen instant the total
radiation energy will be found in the portion $v$ of the volume
$V$ is
%\begin{eqnarray*}
$W = (v/V)^{E/h\nu}$.''\\[6pt]
%$\end{eqnarray*}
\nx
From this he drew the profound conclusion\\[6pt]
\nx
``$\cdots$ monochromatic radiation of low density (within the range
of validity of Wien's radiation formula) behaves
thermodynamically as if it consisted of mutually independent
energy quanta of magnitude $h\nu$.''\\[6pt]
\nx
And then he continued,\\[6pt]
\nx
``$\cdots$ it seems reasonable to investigate whether the laws
governing the emission and transformation of light are also
constructed as if light consisted of such energy quanta.''\\[6pt]
\nx
This is how the concept of the {\bf photon} was born in 1905, though
the name itself was coined much later (in 1926) by the chemist
G. N. Lewis.  Einstein then applied the idea to three known
phenomena. One of these was the photoelectric effect,
discovered by Heinrich Hertz in 1887. Hertz found that if two metal
surfaces are held at a high potential difference, light from a
primary spark on one surface falling on the other surface leads
to another spark.  In 1899 J. J. Thomson showed that when light
falls on a metal surface, the objects liberated are electrons.
In 1902 Philip Lenard discovered that the energy of these
electrons was independent of the light intensity, and found
qualitative evidence for an increase in this energy
with increasing frequency.

There were thus three features of the photoelectric effect
that were not consistent with the wave picture of light.
First, the energy transferred by the light to
the electron does not depend on the light intensity, which
is contrary to expectation because the energy of a wave is
proportional to its intensity. Second, the
frequency of a wave gives the number of disturbances per
unit time. One would therefore expect that a light wave with
a higher frequency (and the same intensity) would liberate more
electrons, but their energy would not increase. This, too,
is not what is observed. Finally,
experiments showed that incident light of a frequency lower than
a threshold frequency (which depended on the metal)
did not liberate any electrons, no matter how much the intensity (and
hence energy) of
the incident radiation was increased. This was puzzling because,
even if one assumed that
there was a threshold or energy barrier
that the electrons had to
overcome to be liberated from the metal,
one would expect that increasing the intensity of
the light would give an adequate impetus to the electrons.
Why should the {\it frequency} of the light be involved?

Einstein answered all these questions in his 1905 paper. He
used the idea of the light quantum to propose the extremely
simple equation
\[E=h\nu - P\]
for the kinetic energy $E$ of the
photo-electron. $P$ denotes the
work function or the energy used up in liberating the electron
from the metal surface.  This equation immediately explained the
apparently strange experimental results,
since the energy of each ``bundle of radiation''
(which produces the photo-electron) is proportional to its
frequency. Increasing the light intensity increases the
number of radiation quanta, and thus increases the number of
photo-electrons, but not the energy of each light quantum.

Today, the equation above is taught in high school, and it
seems so obviously correct --- in hindsight! At the time
Einstein proposed it, however, it was a truly revolutionary idea
that required physicists to give up their well-entrenched
ideas on the wave nature of light. It is therefore not
surprising that considerable
opposition to Einstein's idea persisted
for almost two decades after it was first mooted.
R. A. Millikan did extensive
experiments from 1905 to 1915 and then said,\\[6pt]
\nx
``I spent ten years of my life testing that 1905 equation of
Einstein's, and, contrary to all my expectations, I was
compelled in 1915 to assert its unambiguous verification in
spite of its unreasonableness, since it seemed to violate
everything we knew about the interference of light.''\\[6pt]

In the meantime, Einstein sharpened his concept of the light
quantum.  In 1909 he analyzed the energy fluctuations for
temperature radiation described by the complete Planck Law
(not just the Wien limit), and found that  it was the sum of
two contributions --- one corresponding to a pure Wien Law, and
the other to a pure classical Rayleigh-Jeans Law.  He then
described the Wien contribution in these words:\\[6pt]
\nx
``If it alone were present, it would result in fluctuations
(to be expected) if radiation were to consist of independently
moving point-like quanta with energy $h\nu$.''

Around this time, Einstein took yet another revolutionary step.
He argued that the Planck notion of quantization was not
restricted to light waves alone, but could be extended to
oscillations of other kinds. He was motivated by the fact
that, similar to the breakdown of classical theory is
explaining the blackbody spectrum, there was difficulty in
explaining the low-temperature behaviour of the specific
heat of solids. In 1907, Einstein suggested that one should
treat a crystalline solid as a set of harmonic oscillators of a given
frequency, and calculate its  average (or internal) energy
at a temperature $T$
by assuming that these oscillators had only the discrete
energies proposed by Planck, i.e., that the energy of an oscillator
was related
to its frequency by  $E=nh\nu$. The title of his paper,
``{\it The Planck theory of radiation and the
theory of specific heat''}, says it all. This was the bold first step
toward the correct explanation of the specific heat of
solids, and the first time that the notion of quantization
was applied to oscillations other than light. Although
the complete explanation of the specific heat came from
Peter Debye a few years later, Einstein was one of the first
physicists to accept the idea of quantization as a general
principle.

Later, in 1916, Einstein showed that, besides carrying an energy
$h\nu$, the light quantum also carries a linear
momentum of magnitude $p=h\nu/c$, directed along its direction of
propagation.  After this he wrote in 1917 to his
close friend Michele Besso,\\[6pt]
\nx
``I do not doubt anymore the {\it reality} of radiation
quanta, although I still stand quite alone in this
conviction.''\\[6pt]
\nx
This reflected prevailing continued opposition to the
idea of light quanta --- not only from Millikan, but also ---
surprisingly enough ---
from Planck and Bohr.  The reason was the strong belief that the
phenomena of
interference and diffraction of light implied that the classical
Maxwell wave theory had to be the correct description of
radiation. Quantum effects had therefore
to be limited to matter and
its interaction with radiation. In their nomination of
Einstein for an academic position in Berlin in 1914, Planck, Nernst,
Rubens and Warburg went so far as to add,\\[6pt]
\nx
``That he may sometimes have missed the target in his
speculations, as for example, in his theory of light quanta,
cannot really be held against him.''\\[6pt]
As good an example of ``famous last words'' as any! Even
later, in 1923, Bohr went to the extent of proposing that energy
conservation in individual microscopic events be given up, in
order to save Maxwell's classical description of radiation. But this
was a possibility that Einstein had already considered --- and
rejected,  as early as in 1910.

The final widespread acceptance of the photon idea came only in
1925, after A. H. Compton and A. W. Simon verified the conservation of
energy and momentum in the
Compton effect, that is, in direct photon-electron collisions.

\section*{Brownian Motion}

When microscopic, micron-sized particles such
as pollen grains are suspended in a liquid, they show erratic
and sudden movements as though they were being kicked around in
a random fashion.  This `Brownian motion' is named after the
botanist Robert Brown, who studied it systematically in
1827-28, but the phenomenon was known even earlier.  It had been
thought by some that these irregular and jerky movements were
evidence for `vitalism', a kind of `life-force'.
But after Brown's studies
it became clear that no `vital forces' were involved.  By the
1850's the motion was believed to be caused either by internal motions in
the fluid, or by collisions with fluid molecules from different
directions. Einstein was apparently not too familiar with
the precise details of
earlier experimental work --- or rather, he characterized this work as
too imprecise to enable unambiguous conclusions to be drawn.
This is essentially why the phrase ``Brownian motion'' does not appear
in the title of his first paper on the subject (see Box 3),
although in the text of that paper he says,\\[6pt]
\nx
``It is possible that the motions to be discussed here are
identical with so-called Brownian molecular motion $\cdots$''\\[6pt]
\nx
His aim was far more fundamental:
to show that, if the predictions of the theory
could be experimentally verified, then\\[6pt]
\nx
``$\cdots$ an exact determination of actual atomic sizes becomes
possible.''\\[6pt]
\nx
Indeed, the determination of atomic sizes and of Avogadro's number
$N_{A}$ are
recurring themes in Einstein's early work on statistical physics. He
returned to the determination of $N_{A}$ again and again, proposing
several independent methods to estimate this fundamental quantity. It
is clear that one of his motivations was to establish beyond all doubt
the atomic nature of matter.

Einstein's analysis of Brownian motion was nothing less than
ingenious. Using essentially physical arguments,
he threaded his way through carefully, avoiding pitfalls arising from
what we now know are mathematical subtleties in the behaviour of
certain random processes.
A year before
A. A. Markov introduced
what are now called Markov processes in the theory of probability,
Einstein had essentially recognized
that Brownian motion was a special kind of
Markov process, called a diffusion process.
He correctly identified the distinct
time scales in the problem of a micron-sized object being buffeted
incessantly and randomly by
much smaller molecules, and this helped him write down the equation
governing the probability distribution of (any component of)
the position of the larger
particle, in the form
\[\frac{\partial p(x\,,\,t)}{\partial t}
= D\,\frac{\partial^2 p(x\,,\,t)}{\partial x^2}.\]
This is the famous diffusion equation (also called the heat
conduction equation, as the two are mathematically identical
equations), $D$ being the diffusion coefficient. Einstein
also wrote down the fundamental Gaussian
solution to this equation. If the particle is taken to
start from the origin $x = 0$ at $t = 0$, this solution is
\[p(x\,,\,t) = \frac{1}{\sqrt{4\pi D t}}\,
e^{-x^2/4Dt}\]
for any $t > 0$.
Once these results were in
place, the crucial characteristic feature
of the diffusive process
emerged automatically ---
namely, that the average value of the
{\it square} of the distance travelled in any
given direction by a Brownian particle
in a time interval $t$ is proportional to $t$, rather than $t^2$:
\[
\langle x^2 (t) \rangle =
2Dt, \]
where $D$ is the diffusion coefficient.

Einstein's deep
insight lay in the fact that he concentrated on the mean
squared displacement, rather than the instantaneous velocity
of the particle, as the quantity to be studied and measured.
This is also related to the mathematical
subtleties referred to earlier (see Box 4). He used an
``extremely ingenious'' argument that combined
thermodynamics with dynamics, to relate $D$ to the temperature $T$
of the liquid and its viscosity $\eta$ according to
\[D =
\frac{RT}{6\pi N_A \eta a}\]
where $R$ is the gas constant,
for the case of spherical particles of radius $a$. Therefore
\[
\langle x^2 (t)\rangle =
\frac{RT}{3\pi N_A \eta a}\,t.\]
This makes it possible to determine $N_A$ by a measurement of the mean
square displacement of a Brownian particle over different intervals of
time.

The predictions of Einstein's theory were checked by Jean
Perrin and his students in a series of experiments from
1908 to 1914, and they were all confirmed with ``an until
then unmatched precision''.
With this successful explanation of Brownian motion,
resistance to the reality of atoms (almost!) ended.
Ostwald acknowledged this in 1908, but while Mach also
did so initially, he reverted later to his doubtful
attitude and remained unconvinced till the end.

The Polish physicist Marian von Smoluchowski and the French physicist
Paul Langevin also did pioneering and
extremely significant work on
the problem of Brownian motion and related matters concerning
deep issues such as macroscopic irreversibility,
around the same time as Einstein. Brownian motion has become a paradigm
for a kind of random motion with a staggering variety of applications
 --- for instance, in
stock market fluctuations, dynamic friction in star
clusters in galaxies, and the dynamics of sand-piles, to name just
three of these. The ramifications of Brownian motion
in unexpected areas of mathematics and physics are equally astounding
--- the Gaussian solution
written down above leads, via the so-called Wiener measure,
to the Feynman or path integral formulation of
quantum mechanics, and then on to the modern method of
quantization in quantum field theory.

\section*{The birth of Special Relativity}

Einstein's work on the light quantum and on Brownian motion were
rooted in specific physical phenomena and problems. So was his work on
relativity --- in particular, Special Relativity: it sprang from the
search for a consistent way to describe the electrodynamics of moving
charges, which involves the dynamics of both
material particles and radiation in interaction with each
other. However, once formulated, the principle and postulate of
special relativity  transcend specific phenomena. They
lead directly to deep insights into the nature of space-time itself,
and into fundamental issues such as the symmetry, form-invariance
and observer-independence of physical laws.

To appreciate Einstein's achievement in this regard,
we have to go somewhat further back in history.  Newton's magnum
opus, {\it Philosophiae Naturalis Principia Mathematica}
(the {\it Principia}, as it is generally known), was first published
 in 1687.  In this great book he gave
expression to definite views on the natures of space and time ---
the pre-existing background or arena in
which all natural phenomena
occur.  Essentially, space and time were regarded as individually
absolute and the same for all observers.  Of course inertial
observers and their frames of reference played a distinguished
role, and in them Newton's Laws of mechanics and universal
gravitation are obeyed.

Almost two hundred years later, in 1865, Maxwell presented his
system of equations which unified electricity, magnetism and
optics (the first grand example of unification!)
Light was shown to be a propagating electromagnetic wave,
with a speed calculable from electric and magnetic measurements.
It soon became clear that there was a clash between Newton's
treatment of space and time, and the Maxwell theory. The speed of
light in a vacuum (or free space)
could be as predicted only in a sub-class of the Newtonian
inertial frames, all of which would have to be at rest with
respect to each other.
In all other inertial frames, this speed
would have to be
variable, dependent on the motion of the observer. However,
all attempts to detect this frame-dependence of the speed of light
failed.
The most famous experiments were carried out in 1887 by Michelson
and Morley, working at the Case School of Applied Science and
Western Reserve University in Cleveland, Ohio. These
experiments thus showed that Maxwell was correct, not Newton.

Many scientists attempted to reconcile Newtonian mechanics with
the Maxwell theory, the most prominent being Lorentz, Fitzgerald
and Poincar\'{e}.  But their efforts were unconvincing, and ultimately
unsuccessful. The definitive answer came with
Einstein's work in 1905, where he
re-analyzed the nature of space and time.  They are not
individually absolute and the same for everybody, as Newton had
visualized; rather, it is only the combined space-time continuum
which is common to all, but each inertial observer divides it up into a
space and a time in her own way.  The difference can be
illustrated in the following manner. Imagine two events which
occur at two different spatial locations at two unequal times.
Comparing the observations of two different inertial
observers of these two events, one finds the following distinction
between the old (or Newtonian, non-relativistic) description, and
the new (or Einsteinian, relativistic) description:
\begin{eqnarray*}
\begin{array}{lcccccc} &&&\underline{\mbox{Spatial separation}}&&&
\underline{\mbox{Time
separation}}\nonumber\\
&&&&&&\nonumber\\
\mbox{Newtonian view}&&&\mbox{different}&&&\mbox{same}\nonumber\\[6pt]
\mbox{Einsteinian
view}&&&\mbox{different}&&&\mbox{different}\nonumber\\
&&&&&&\nonumber\\
\end{array}\end{eqnarray*}
In essence, the simultaneity of spatially
separated events, and the time interval between events,
are not absolute concepts. They
are both dependent
on (the state of motion of) the observer.  Thus Einstein's resolution
of the conflict was to modify Newtonian mechanics while retaining
Maxwell's theory --- the former had to fall in line with the latter.
Later in 1905 he obtained, as a consequence of the modified
mechanics, the famous formula $E=mc^2$.  Special relativity was
thus found via Maxwell's theory of electromagnetism.  But we must
also recognise that Einstein was already aware that this
classical Maxwell theory
itself was in need of modification, as was indicated by the
failure of the Rayleigh-Jeans
law for temperature radiation, and the evidence for the
quantum nature of light.

As we have already mentioned,
Special Relativity is really a basic principle
applicable to all of
physics (except gravitation)!
Here are two expressions of this idea:\\[6pt]
\nx
From a lecture by Einstein in 1911 ---
``The Principle of Relativity is a principle that narrows the
possibilities; it is not a model, just as the Second Law of
Thermodynamics is not a model.''\\[6pt]
\nx
And from a review by V.Bargmann ---
``$\cdots$ every physical theory is supposed to conform to the basic
relativistic principles and any concrete problem involves a
synthesis of relativity and some specific physical theory.''\\[6pt]
\nx
Examples of this are the Dirac equation for the electron, the theory
of quantum electrodynamics and the subsequent unified electroweak theory,
and the currently accepted quantum chromodynamic theory of strong
interactions --- in fact, the entire standard model of elementary
particle physics, which is ultimately
{\it all} of fundamental physics except
for gravitation.

\section*{Life after 1905}
It is the centenary of these remarkable
achievements of Einstein in 1905 that are being
celebrated this year throughout the world.  Any one of these three
pieces of work by a single person would have established that
individual's reputation for life.
What is awesome is that Einstein did all three of them (see Box 5).
As Abraham Pais says in his definitive biography of Einstein, ``No one
before or since has widened the horizons of physics in so short a time
as Einstein did in 1905.''

To round off the picture, let us recount briefly
some of the significant later developments in
physics in which Einstein played the leading role or to which he
 contributed
in significant measure.\\

\nx
{\bf 1909}:\, As we mentioned earlier, by using the
complete Planck Law Einstein showed that the energy fluctuations
of temperature radiation are the sum of two terms --- a non-classical
particle like Wien term, and a classical wave like Rayleigh-Jeans
term.  Einstein described their simultaneous presence thus:\\[6pt]
\nx
``It is my opinion that the next phase in the development of
theoretical physics will bring us a theory of light that can be
interpreted as a kind of fusion of the wave and the emission theory
$\cdots$
{\it  The wave structure and the quantum structure are not to be
considered as mutually incompatible}'' [emphasis added].\\[6pt]
\nx
Thus, this was the first clear recognition of {\bf wave-particle
duality} in physics.\\

\nx {\bf 1907-1915}:\,  During this decade Einstein
steadily built up his General Theory of Relativity.  In attempting
to bring together Newton's theory of gravitation and Special
Relativity, he saw that it was necessary to supersede both of
them. Gravity found a new interpretation as curvature of
space-time, and geometry became a dynamical entity, a part of
physics influenced by, and influencing, the rest of nature.  It
should be emphasized that, while Special Relativity amounted to a
requirement on all of physics except gravitation, General
Relativity is the final classical theory of gravitation itself,
with rules for determining the effects of gravity on all other
interactions. It is, to quote Landau and Lifshitz, ``the classical
field theory {\it par excellence}''.\\

\nx
{\bf 1916}:\,  Planck's Law appeared in Einstein's work
many times --- in 1905, in 1909, then again in 1916 when he gave a
startlingly new derivation of it based on Bohr's idea of discrete
stationary states of atoms, and spontaneous and stimulated emission
and absorption of radiation by matter.  Already in Rutherford's
exponential law for radioactive decay in 1900, the notion of
probability had come into physics in an important way, apart from
its use in statistical mechanics.  Through his work Einstein showed
that  this mathematical concept played a role at the most
fundamental level in the atomic domain.  Almost four decades later,
the concept of stimulated emission was exploited in the development
of the maser and the laser (see Box 6).\\

\nx
{\bf 1917}:\,  This year saw Einstein applying general
relativity to the field of cosmology, but it turned out to be
somewhat premature, as Hubble's discovery of other galaxies and
the expansion of the
universe was still some twelve years away.\\

\nx
{\bf 1925}:\,  Building on the discovery of Bose
statistics by Satyendra Nath Bose in 1924,
Einstein gave the first theory of
the ideal quantum (or Bose) gas\footnote{
See N. Mukunda, {\it Bose Statistics -- Before and After},
Current Science {\bf 66}, 954-964 (1994).},
and predicted the phenomenon that
has become known
as Bose-Einstein condensation. Parallel to the 1909 energy
fluctuation formula for radiation, he now  obtained a density
fluctuation formula for the material quantum gas --- it appeared
now as the sum of a non-classical wave term and a classical particle
term. This meant that matter too had to exhibit
wave-particle duality.\\

\nx
{\bf 1925-1927}:\,  This two-year period saw the
creation of quantum mechanics by Werner Heisenberg,
Erwin Schr\"odinger and
P. A. M. Dirac. It also witnessed the emergence of the
so-called orthodox or Copenhagen interpretation with inputs from
many, including  Born, Bohr, Heisenberg, Jordan and
Pauli.  Heisenberg's Uncertainty
Principles and Bohr's Complementarity Principle formed important
components of this interpretation.  At crucial stages both
Heisenberg and Schr\"odinger drew inspiration from conversations
with and remarks by Einstein. However, even though he had done so
much to prepare the ground for the advent
of quantum mechanics, Einstein
never accepted the orthodox interpretation or the claim of the
finality of quantum mechanics.\\

\nx
{\bf 1927-1930}:\,  Initially, Einstein tried to show
that quantum mechanics was incorrect, by devising subtle
experimental arrangements which could circumvent the uncertainty
principles.  This happened with respect to the position-momentum
uncertainty principle during the 1927 Solvay Conference, and the
time-energy uncertainty principle at the next Solvay
Conference, in 1930.  However, on both occasions Bohr was
able to counter
Einstein's arguments and prove the consistency of quantum mechanics.
Einstein accepted Bohr's replies, but remained unconvinced of the
finality of quantum mechanics. \\

\nx
{\bf 1935}:\,  Einstein then changed his stand, and in
a landmark paper with Boris Podolsky and Nathan Rosen he argued
that, while quantum mechanics may well be internally consistent, it
was {\it incomplete}.  They proposed retaining what they called locality
and realism in any complete physical theory, both of which are
violated by standard quantum mechanics.  The most important effect
of their paper has been to highlight a key feature of quantum
mechanics called {\bf entanglement}.  In fact, in an important
contribution by Schr\"odinger within the year, this term was
introduced for the first time; and Schr\"odinger went so far as to
say that this was {\it the} key feature, not {\it one}
of the features, of quantum mechanics.  In picturesque language
the idea can be conveyed thus: in the quantum mechanics of
composite systems, the whole can be greater than the sum of the
parts, as the latter cannot capture subtle quantum correlations.
Over the decades, experiments of increasing sensitivity have ruled
in favour of quantum mechanics and against the Einstein point of
view.  Today quantum entanglement is referred to as a resource or
currency for carrying out quantum computation.\\

To conclude, the importance of Einstein's work in 1905 for later
developments in physics is amply evident.  Usually, advances in
physics, or indeed in any part of science,
take place in a more-or-less
steady and cumulative manner. Each step forward is built on a chain of
earlier advances, and is rarely an isolated breakthrough.
Occasionally, however,
there occur major advances, steps into stunningly
new ways of
thinking ({\it paradigm shifts}, in fashionable language),
which completely alter the landscape of the subject.
This happened with each of Einstein's achievements in 1905.  It
happened again with Niels Bohr's atomic model in 1913, with
General Relativity in 1915, and with the advent of quantum
mechanics in 1925-27.  Cause enough for celebration!

\section*{Suggested Reading}
\begin{itemize}{}{}

\item[1.] \emph{`Albert Einstein: Philosopher-Scientist', The Library of Living Philosophers}, Vol.VII, ed. P A Schilpp, Open Court, La Salle, Illinois, 1949.

\item[2.] A Pais, \emph{`Subtle is the Lord \ldots' The Science and the Life of Albert Einstein}, Oxford University Press, Oxford, 1982.

\item[3.] J Stachel, \emph{Einstein's Miraculous Year: Five Papers that Changed the Face of Physics}, Princeton University Press, 1998.

\item[4.] J S Rigden, \emph{Einstein 1905: The Standard of Greatness}, Harvard University Press, Cambridge, Mass., 2005.\\

{\bf In addition, the reader may consult the following articles} \\

\item[5.] N Mukunda, Bose Statistics -- Before and After,
\emph{Current Science}, Vol. 66, pp. 954--964, 1994.

\item[6.] S R Madhu Rao, Special Relativity -- An Exoteric Narrative, \emph{Resonance}, Vol. 3, No. 1, pp. 61--66, 1998.

\item[7.] S R Madhu Rao, Special Relativity -- An Exoteric Narrative, \emph{Resonance}, Vol. 3, No. 5, pp. 63--72, 1998.

\item[8.] John Stachel, Albert Einstein -- The Man Behind the Myths, \emph{Resonance}, Vol. 3, No. 8, pp. 76--92, 1998.

\item[9.] \emph{Resonance}, Vol. 5, Nos.3 and 4, 2000, Special Einstein Issues.

\item[10.] Vasant Natarajan, Einstein as Armchair Detective: The case of Stimulated Radiation, \emph{Resonance}, Vol. 6, No. 6, pp. 28--42, 2001.

\end{itemize}

\newpage

\begin{center}
{\bf Box 1}
\end{center}
The three seminal papers  published by Einstein in his
``miraculous year'' in Annalen der
Physik are, in chronological order:
\begin{enumerate}
\item[(i)]\, {\it On a heuristic point of view concerning the
production and transformation of light}, Vol. {\bf 17},
pp.  132-148. Received March 18, 1905.

\item[(ii)] \,{\it On the motion required by the
molecular kinetic theory of heat of particles suspended in fluids at rest},
Vol. {\bf 17}, pp. 549-560. Received May 11,
1905.

\item[(iii)]\,{\it On the electrodynamics of moving bodies}, Vol.
{\bf 17}, pp. 891-921. Received June 30, 1905.
\end{enumerate}
\nx
The relation $E = mc^2$ appeared for the first time in
\begin{enumerate}
\item[(iv)]\,{\it  Does the inertia of a body depend upon its energy
content?}, Vol. {\bf 18}, pp. 639-641. Received
September 27, 1905.
\end{enumerate}

% \newpage

\begin{center}
{\bf Box 2}
\end{center}
It is difficult, if not impossible,
to make a fair comparison of
truly outstanding achievements in any field of human endeavour if
these are widely separated in time and circumstance.
(Is the greatest batsman to date
Bradman or Tendulkar?) And yet human interest in {\it
  records} and extrema is insatiable. What would qualify as
the {\it most}
intense and sustained mental effort by a single person leading to the
most profound results? Newton, Gauss, Darwin and Einstein, each at the
peak of his creative outburst, would
surely qualify to be very near, if not at the top, of this exclusive
list. Clearly, proper mental preparation was an essential condition ---
their minds had to be
congenial receptacles and fertile ground
for the new ideas to germinate and grow. And
each of these great figures did indeed ``stand on the shoulders of
giants'' who preceded
them, to see further. For, in Science, {\it there is no room for
any miraculous revelation --- or for unquestionable dogma},
for that matter.

% \newpage

\begin{center}
{\bf Box 3}
\end{center}
The title of Einstein's first paper on Brownian motion was
{\it On the motion required by the molecular kinetic theory of heat of
particles suspended in fluids at rest}. This paper was received by
{\it Annalen der Physik} just eleven days after Einstein's doctoral
thesis was completed, although the thesis itself was only published
in 1906. The thesis contains results quite as fundamental as those
Einstein published in his {\it Annus Mirabilis}. In fact, the
marvelous formula relating the diffusion coefficient, Avogadro's
number, viscosity and the temperature appeared there for the first
time. His second paper on Brownian motion, in December 1905,
 gets right to the
point, being titled simply
{\it On the theory of Brownian motion}.

% \newpage

\begin{center}
{\bf Box 4}
\end{center}
Here are some of the peculiarities of the `sample path' of a particle
undergoing Brownian motion in the strict mathematical sense. Its
instantaneous
velocity turns out to be unbounded. Its trajectory  is
a continuous,
but extremely jagged, curve. It is an example of a {\it random
fractal}: it
is {\it non-differentiable} almost everywhere, and is said to be
{\it statistically self-similar}.
That is, its degree of jaggedness remains
unchanged under arbitrarily large magnification of any portion of the
curve. The curve is {\it space-filling}, in the following sense: if
the Brownian motion is restricted to an infinite
line or an infinite plane, then every
point of the line or plane is sure to be visited infinitely often by
the particle as $t \to \infty$. However, the
mean time between successive visits is infinite.
If the Brownian motion occurs in
three-dimensional space, the so-called {\it fractal dimension}
of its trail is $2$, and not $1$ as would be expected of an
ordinary regular curve.

% \newpage

\begin{center}
{\bf Box 5}
\end{center}
The urge to compare being an irrepressible human quality, one is
tempted to ask: which of Einstein's stupendous achievements is his
{\it greatest} contribution to physics, at least in hindsight? An
extremely difficult question, given the awe-inspiring {\it depth} of his
insight. An excellent case can be made out in
favour of his contributions to each one of the major subjects he
tackled: statistical physics, quantum physics, relativity and
gravitation. Some underlying themes can be distinguished. To
list a few of these, he
had the most profound insight into the
fundamental role of {\it fluctuations,  symmetry, invariance,
causality}, and into the {\it non-locality inherent in
quantum mechanics}.

In order to give an illustration of the way Einstein
thought about physical
problems, and the manner in which he combined physical arguments
to arrive at far-reaching results, we summarize in Box 6 a specific
instance, namely, his work on the stimulated
emission of radiation. This
led, when the technology became available, straight to the laser.

% \newpage

\begin{center}
{\bf Box 6}
\end{center}
Einstein's 1916 paper, titled  {\it On the quantum theory of
radiation}, is a {\it tour de force} in  physics.
Using simple arguments,\footnote{
Einstein's approach in this paper
represents a kind of {\it modus operandi} for much of
his work (except, perhaps, General Relativity).
For a detailed analysis of the paper,
see Vasant Natarajan, {\it Einstein as armchair detective: The case
of stimulated emission} in {\it Resonance} {\bf 6}, No. 6, pp. 28-42
(June 2001).}
he was able to predict several new
features of the interaction between matter and radiation:
the process of stimulated emission; the relation between the
coefficients for emission and absorption (the Einstein $A$
and $B$ coefficients, still used in modern terminology); and
the discrete momentum $h\nu/c$ carried by each photon.

He starts the paper with the profound statement, ``The
formal similarity between the chromatic distribution curve
for thermal radiation and the Maxwell velocity-distribution
law is too striking to have remained hidden for long''.
With this motivation, he proceeds to understand the
features of matter-radiation interaction from the point of
view of thermodynamic equilibrium.
The year is 1916. He is therefore quite familiar with the Planck
hypothesis of radiation quanta, having used it to explain
the photoelectric effect; he is aware of the Bohr model to
explain the discrete nature of atomic spectra; and he is of course
a master at using thermodynamic arguments, right from his
doctoral thesis work on Brownian motion. But
quantum mechanics itself, or the Schr\"odinger equation, is not yet in
place. Still,
Einstein is able to predict many new `quantum'
features of radiation.

Einstein considers
a gas of atoms at a temperature $T$ and assumes that
each atom has only two energy levels. He then makes
certain hypotheses about the processes of absorption and
emission of radiation for transitions between these levels.
He then requires that, under thermal equilibrium, the rate
of absorption should be balanced by the rate of emission,
so that the equilibrium occupancy of the two levels remains
unchanged. He shows that this is possible only if one
postulates the new process of stimulated emission, in
addition to the known process of spontaneous emission. With
this process included, he is able to give a simple, new
derivation of Planck's radiation formula, and further show
that the frequency of the emitted radiation is related to
the difference in the atomic energy levels
by the Bohr principle, $\Delta E =
h\nu$. Going further, he states
that the exchange of momentum between the atoms and radiation
(and the consequent change in velocity of the atoms) should
not affect the thermal (Maxwell) velocity distribution. He
now uses his deep insight into Brownian motion (this time
in momentum space) to show that this is possible only if
each ``radiation bundle'' carries a momentum $h\nu/c$ along
its direction of propagation.

Einstein's prediction of stimulated emission led, almost
forty years later, to the development of the maser and the
laser. Today lasers are found everywhere: in your
computer's CD-ROM drive, in the grocery-store scanner, in
the doctor's office, in fibre-optic telecommunications, and
in research laboratories. The momentum carried by photons
demonstrated in this paper leads to radiation
pressure, which is important in situations ranging from
isotope separation to laser cooling of atoms. And
stimulated emission in a more general {\it avatar}, called
stimulated scattering of bosons, leads to the phenomenon of
Bose-Einstein condensation of a gas, as first shown by
Einstein in 1924. This new state of matter was
experimentally created in the laboratory in 1995.

To gauge the impact of this paper by Einstein,
note that no fewer than
four Nobel Prizes in Physics have been awarded in recent times for
related developments: in 1964 (laser/maser action), in 1981 (laser
spectroscopy), in 1997 (laser cooling and trapping), and in 2001
(Bose-Einstein condensation).

\section*{Authors introduction:}

Vasant Natarajan is at
the Department of Physics,
IISc, Bangalore. His current
research involves trapping of
atoms to carry out high
precision tests of fundamental
physics. He has earlier
worked on high precision
mass spectrometry and on the
focussing of atomic beams by
laser fields. \\

\nx
V Balakrishnan is at
the Department of Physics,
IIT- Madras, Chennai. His
research interests are in
dynamical systems,
stochastics and statistical
physics. \\

\nx
N Mukunda is at
the Centre for High Energy
Physics, IISc, Bangalore. His
interests are classical and
quantum mechanics,
theoretical optics and
mathematical physics.

\end{document}